\documentclass[aps,floatfix,amsmath,nofootinbib,amssymb,towcolumn,superscriptaddress]{revtex4}

\usepackage{overpic}
\usepackage{amssymb}
\usepackage{indentfirst}
\usepackage{feynmf}   
\usepackage{slashed}  
\usepackage{cases}
\usepackage{color}
\usepackage{multirow}
\usepackage{epstopdf}
\usepackage{graphicx,color,bm}
\usepackage{epstopdf}

\usepackage[colorlinks,
            citecolor=green,
            anchorcolor=red,
            menucolor=red,
            linkcolor=red,
            filecolor=red,
            runcolor=red,
            urlcolor=blue,
            frenchlinks=red]{hyperref}

\begin{document}

\title{Quark spin-orbit correlations in the pion meson in light-cone quark model}

\author{Chentao Tan}\affiliation{School of Physics, Southeast University, Nanjing
211189, China}

\author{Zhun Lu}
\email{zhunlu@seu.edu.cn}
\affiliation{School of Physics, Southeast University, Nanjing 211189, China}

\begin{abstract}

We study the correlation between the quark spin and orbital angular momentum inside the pion meson. Similar to the case inside the nucleon, the longitudinal spin-orbit correlation $C_z^{q/\pi}$ in pion meson can be expressed in terms of the corresponding generalized parton distributions (GPDs) and generalized transverse momentum distributions (GTMDs). This provides new information about the spin structure of the pion. Using the wavefunctions of the pion in the light-cone quark model and the overlap representation for GPDs and GTMDs, we present the analytical results for the quark longitudinal spin-orbit correlation. We find that the GPD approach and the GTMD approach lead to the same results. The numerical results is also obtained, showing that the correlation in pion is anti-aligned. In addition, we compare $C_z^{q/\pi}$ from the GPD approach and the GTMD approach, with $x$ and the transverse momentum $k_T$ unintegrated.
\end{abstract}

\maketitle

\section{Introduction}

Understanding the spin content of hadrons has been recognized as one of the main goals in hadronic physics~\cite{Jaffe:1989jz,Ji:1996ek,Leader:2013jra}. Particularly, the correlations between the parton/hadron spin and the orbital motion of partons inside hadron can bring much broader contents to the spin and partonic structure of hadrons.
For example, the correlation between the transverse spin of the nucleon and the parton transverse momentum
leads to the a novel distribution called as Sivers function~\cite{Sivers:1989cc,Sivers:1990fh,Brodsky:2002cx}, which is the asymmetric distribution of the unpolarized parton in the transversely polarized nucleon.
Recently, the parton longitudinal spin-orbit correlation~\cite{Lorce:2014mxa}, sketched by $\langle L_z^a S_z^a\rangle$, has also received a lot of attentions.
It describes the difference between the right-handed and left-hand quark contributions to the quark longitudinal orbital angular momentum (OAM), and provides a new piece of independent information about the longitudinal spin structure of hadrons.
Another advantage of the parton longitudinal spin-orbit correlation is that this correlation is invariant under the parity transformation.

In Ref.~\cite{Lorce:2014mxa}, the parton spin-orbit correlation in the nucleon has been studied in details. In particularly, a local gauge-invariant operator definition for the longitudinal spin-orbit correlation is reported, and the quantitative relations between the quark spin-orbit correlation and the moments of the twist-2 or twist-3 (GPDs are provided. In this way, the spin-orbit correlation can be accessed through measurable observables. These extend the previous study~\cite{Lorce:2011kd} that the information of spin-orbit correlations can be deduced from the GTMDs~\cite{Meissner:2008ay,Meissner:2009ww,Lorce:2013pza}, which are difficult to measure experimentally so far.

In this work, we study the correlation between the longitudinal spin and the quark orbital angular momentum of valence quarks inside the pion.
As the pion meson is a spin-0 hadron, the knowledge of its spin structure in terms of polarized partons is less known and seldom investigated.
Fortunately, the parton longitudinal spin-orbit correlation does not require polarization of a hadron. Therefore, in principle one can explore this effect inside spin-0 hadron such as pion meson. For this purpose we apply a light-cone quark model to provide relations for the spin-orbit correlation of pion meson in terms of pion GPDs or GTMDs.
It is also interesting to verify if the GPD approach and the GTMD approach can lead to the same result for spin-orbit correlation.
We will clarify this result for the pion as a case study.

The rest of the paper is organized as follows: In section II, we define the quark spin-orbit correlation operator in the pion and express the corresponding expectation value in terms of form factors. In section III, we relate the form factors with specific moments of the pion GPDs/GTMDs. In section IV, we provide the analytic results as well as the numerical results of the correlation using the pion wavefunctions deduced from a light-cone quark model. We summarize the paper in Section V.

\section{Definition}

The gauge-invariant light-front operator for quark longitudinal OAM has attracted a lot of interests because it enters the Ji decomposition of the longitudinal spin of the nucleon~\cite{Ji:1996ek}
\begin{align}
\hat{J}_z=\hat{S}^q_z+\hat{L}^q_z+\hat{J}^G_z.
\end{align}
Here, $\hat{L}^q_z$ represents the operator for quark longitudinal OAM, which is the sum of the left- and right-handed quark contributions:
\begin{align}
\hat{L}^q_z&=\int d^3x \frac{1}{2} \bar{\psi} \gamma^+(\bm{x} \times i\overleftrightarrow{\bm{D}})_z \psi
=\hat{L}^{q_R}_z+\hat{L}^{q_L}_z,
\label{eq:z}
\end{align}
where $\overleftrightarrow{\bm{D}}=\overleftarrow{\bm{\partial}}-\overrightarrow{\bm{\partial}}-2ig\bm{A}$ is the symmetric covariant derivative, $\psi_{R,L}=\frac{1}{2}(I\pm \gamma_5)\psi$, $a^{\pm}=\frac{1}{\sqrt{2}}(a^0\pm a^3)$ with $a$ denoting a generic four-vector, $d^3x=dx^-d^2x_{\perp}$.
However, the complete characterization of the spin structure also requires the knowledge of quark spin-orbit correlations.
Particularly, the gauge-invariant longitudinal spin-orbital correlation describes the difference between these left- and right-handed quark contributions~\cite{Lorce:2014mxa}:
\begin{align}
\hat{C}_z^q&=\int d^3x \frac{1}{2} \bar{\psi} \gamma^+ \gamma_5 (\bm{x} \times i\overleftrightarrow{\bm{D}})_z \psi=\hat{L}^{q_R}_z-\hat{L}^{q_L}_z.
\label{eq:w}
\end{align}
This kind of quark spin-orbit correlation inside the spin-1/2 hadron (the nucleon) was studied in Ref.~\cite{Lorce:2014mxa} in details, but has not been investigated in the case of the spin-0 hadron, such as pion meson.
Thus, the study of this effect inside the pion meson will provide unique information on the longitudinal-spin of the quark and orbital motion of quarks inside a spin-0 hadron, which has seldom been explored so far.

The matrix elements of the operator in Eq.~(\ref{eq:w}) can be parameterized in terms of form factors. To do this one can starts from the parametrization of the matrix elements of the energy momentum tensor $\hat{T}_q^{\mu \nu}$, since the quark OAM operator can also be expressed in terms of $\hat{T}_q^{+i}$
\begin{align}
\hat{L}^q_z=\int d^3x(x^1\hat{T}_q^{+2}-x^2\hat{T}_q^{+1}), \label{eq:oam}
\end{align}
with $\hat{T}^{\mu\nu}$ has the form~\cite{Leader:2013jra}
\begin{align}
\hat{T}_q^{\mu\nu}&=\frac{1}{2} \bar{\psi} \gamma^\mu i\overleftrightarrow{D}^\nu \psi\\
                &=\hat{T}^{\mu\nu}_{q_R}+\hat{T}^{\mu\nu}_{q_L},
\end{align}
where $\hat{T}^{\mu\nu}_{q_{R,L}}=\frac{1}{2} \bar{\psi}_{R,L} \gamma^\mu i\overleftrightarrow{D}^\nu \psi_{R,L}$.
Spin-0 hadrons such as the pion meson whose quark energy momentum tensor has been studied can be parameterized in terms of three form factors\cite{Tanaka:2018wea,Freese:2019bhb,Polyakov:2018zvc,Krutov:2020ewr}
\begin{align}
\langle{p^{\prime}} | \hat{T}^{\mu \nu}_{q} (0) | p \rangle=2P^{\mu}P^{\nu}A_q(t) + \frac{1}{2} (\Delta^{\mu}\Delta^{\nu}-g^{\mu\nu}\Delta^2)D_q(t) +2M_\pi^2g^{\mu\nu} \bar{c}_q(t), \label{eq:pi_emt}
\end{align}
where $M_\pi$ is the pion mass, $P=\frac{p^{\prime}+p}{2}$ is the average four-momentum, and $t=\Delta^2$ is the square of the four-momentum transfer $\Delta=p^{\prime}-p$. Substituting Eq.~(\ref{eq:pi_emt}) into Eq.~(\ref{eq:oam}), one finds that $L_q$, the OAM of quarks inside the pion, is actually zero.

Similarly, one can also write the quark spin-orbit correlation operator as~\cite{Lorce:2014mxa}
\begin{align}
\hat{C}^q_z=\int d^3x(x^1\hat{T}_{q5}^{+2}-x^2\hat{T}_{q5}^{+1}),\label{eq:czoperator}
\end{align}
where $\hat{T}_{q5}^{\mu\nu}$ is the parity-odd partner of the quark energy-momentum tensor operator and has the form
\begin{align}
\hat{T}_{q5}^{\mu\nu}&=\frac{1}{2} \bar{\psi} \gamma^\mu \gamma_5 i\overleftrightarrow{D}^\nu \psi\\
                     &=\hat{T}^{\mu\nu}_{q_R}-\hat{T}^{\mu\nu}_{q_L}.
\end{align}

The non-forward matrix element of the parity-odd operator $\hat{T}_{q5}^{\mu\nu}$ sandwiched by two pion states can be parameterized in terms of two form factors~\cite{Lorce:2014mxa}~\footnote{$\tilde{A}$ does not appears here since  $\tilde{A}$ is the hadron-spin dependent form factor, while pion is spin-0.}
\begin{align}
\langle{p^{\prime}} | \hat{T}^{\mu \nu}_{q5}(0) | p \rangle= - {P^{[\mu}i\epsilon^{\nu]+\Delta P}\over 2 P^+}( \tilde{C}_q(t)-2\tilde{F}_q(t))+i
\epsilon^{\mu\nu\Delta P}\tilde{F}_q(t)+\mathcal{O} (\Delta^2),
\label{eq:q}
\end{align}
where $\epsilon^{\mu\nu\alpha\beta}$ is a totally antisymmetric tensor, with $\epsilon^{+-12}=1$.

Substituting Eq.~(\ref{eq:q}) into the matrix element of Eq.~(\ref{eq:czoperator}) and working with $\bm{P}_\perp=\bm{0}_\perp$, which is the case of the light-cone frame:
\begin{align}
C^{q/\pi}_z \equiv \frac{\langle{p} | \hat{C}^q_z | p \rangle}{{\langle{p} | p \rangle}} = \tilde{C}_q(0).
\label{eq:r}
\end{align}
Here, a covariant normalization of pion states has been used: $\langle{p}^\prime | p \rangle=2p^0(2\pi)^3\delta^3(\bm{p}^\prime-\bm{p})$.
Therefore, in order to obtain the correlation in the pion meson, one only needs to measure the form factor $\tilde{C}_q(t)$.

\section{connection between spin-orbit correlation in the pion and GPDs/GTMDs}

As in the case of energy-momentum tensor, there is no fundamental probe that can couple to $\hat{T}^{\mu\nu}_{q5}$.
Therefore, we will re-represent $\hat{T}^{\mu\nu}_{q5}$ by relating the corresponding form factors to the specific moments of the GPDs or GTMDs.
The relation can be obtained using the QCD identity
\begin{align}
\bar{\psi} \gamma^{[\mu}\gamma_{5}i \overleftrightarrow{D}^{\nu]} \psi=2m\bar{\psi} i\sigma^{\mu\nu} \gamma_{5} \psi-\epsilon^{\mu\nu\alpha\beta} \partial_{\alpha} (\bar{\psi} \gamma_{\beta} \psi)\,,
\end{align}
where $m$ is the quark mass.
Taking the non-forward matrix elements of both sides in the above equation, the left-hand side corresponds to the spin-orbit correlation, while we finds for the right side:
\begin{align}
\langle{p^{\prime}} | \bar{\psi} \gamma^{\mu} \psi | p \rangle= \Gamma^{\mu}_{qV}
\end{align}
\begin{align}
\langle{p^{\prime}} | \bar{\psi} i\sigma^{\mu\nu} \gamma_{5} \psi | p \rangle=\Gamma^{\mu\nu}_{qT}
\end{align}
with
\begin{align}
\Gamma^{\mu}_{qV}=2P^{\mu} \int F_1^{q/\pi}(x,\xi,t) dx
\end{align}
\begin{align}
\Gamma^{\mu\nu}_{qT}=\frac {2i\epsilon^{\mu\nu\alpha\beta} \Delta_{\alpha} P_{\beta}} {M_\pi} \int H_1^{q/\pi}(x,\xi,t) dx
\end{align}
where $\xi=-\Delta^+/2P^+$ is the skewness variable, and $F_1^{q/\pi}(x,\xi,t), H_1^{q/\pi}(x,\xi,t)$ are the  twist-2 GPDs~\cite{Meissner:2008ay,Burkardt:2007xm} of the pion parameterizing the non-local axial-vector and tensor light-front quark correlators, respectively
\begin{align}
&{1\over 2}\int{dz^- \over 2\pi} e^{ixP^+ z^-}\langle p^\prime| \bar{\psi}\left(-{z^-\over 2}\right)
\gamma^+ {\psi}\left({z^-\over 2}\right)|p\rangle=F_1^{q/\pi}(x,\xi,t)\\
&{1\over 2}\int{dz^- \over 2\pi} e^{ixP^+ z^-}\langle p^\prime| \bar{\psi}\left(-{z^-\over 2}\right)
 i\sigma^{j+}\gamma_5{\psi}\left({z^-\over 2}\right)|p\rangle =-\frac{i\epsilon_\perp^{ij}\bm \Delta_\perp^i}{M_\pi}H_1^{q/\pi}(x,\xi,t)
\end{align}

Therefore, the spin-orbit correlation can be determined by the combination of the moments of $F_1^{q/\pi}(x,\xi,t)$ and $H_1^{q/\pi}(x,\xi,t)$
\begin{align}
\tilde{C}^{q/\pi}(t)=\int dx\left (\frac{m}{M_\pi} H^{q/\pi}_1(x,\xi,t)- \frac{1}{2} F^{q/\pi}_1(x,\xi,t)\right),
\end{align}
then the expectation value of quark spin-orbit correlation operator is given
\begin{align}
C^{q/\pi}_z= \int dx \left(\frac{m_q}{M_\pi} H^{q/\pi}_1(x,0,0)- \frac{1}{2} F_1^{q/\pi}(x,0,0)\right)
\label{eq:1}
\end{align}
where $\xi,t=0$.
A comparison can be made with the correlation in the nucleon~\cite{Lorce:2014mxa}
\begin{align}
C^{q/n}_z=\frac{1}{2}\int dx x \tilde{H}_q(x,0,0)-\frac{1}{2}\left(F^{q/n}_1(0)-\frac{m}{2M}H_1^{q/n}(0)\right),
\end{align}
where the superscript $q/n$ represents the quark flavor $q$ in nucleon $n$, $\tilde{H}(x,\xi,t)$ is helicity-fip GPD, $F^{q/n}_1(t)$ is the Dirac form factor, and $H^{q/n}_1(t)$ is a tensor form factor.
In Ref.~\cite{Lorce:2014mxa}, in order to estimate $C^{q/n}_z$, the light-front constituent quark model and the light-front chiral quark-soliton model~\cite{Lorce:2011dv} have been applied to calculate the moments of $\tilde{H}_q(x,0,0)$, the results are compared with experimental measurements~\cite{Leader:2010rb} and lattice calculation~\cite{LHPC:2010jcs}.
The main difference between the pion case and the nucleon case is that the helicity-flip $\tilde{H}$ also contributes to the spin-orbit correlation of the nucleon. This is because the $\tilde{H}$  does not exist in the case of the pion meson.

As derived in Refs.~\cite{Lorce:2011kd,Chakrabarti:2016yuw,Kaur:2019kpi}, the spin-orbit correlations can be also expressed in terms of GTMDs. Particularly, $C_z$ is connected to the GTMD $G_{1,1}$ by the relation
\begin{align}
C^{q/\pi}_z=\int dx d^2\bm k_{\perp} \frac{\bm{k}^2_{\perp}}{M^2} G_{1,1}^{q/\pi}(x,0,\bm{k}^2_{\perp},0,0).
\label{eq:cqgtmd}
\end{align}
where $G_{1,1}^\pi(x,\xi,\bm{k}^2_\perp,\bm{k}_\perp \cdot \bm \Delta_\perp,\bm \Delta^2_\perp)$ is defined as
\begin{align}
W^{[\gamma^+ \gamma_5]} & =
 -\frac{i\varepsilon_\perp^{ij} \bm k_\perp^i \bm \Delta_\perp^j}{M_\pi^2} \, G_{1,1}^\pi \,,
\label{eq:g11}
\end{align}
with the notation
\begin{align}
&W^{\Gamma}(x, P, \bm k_\perp,\Delta)  = {1\over 2} \textrm{Tr}
[W(P, x, \bm k_\perp, \Delta)\Gamma]\nonumber\\
& \qquad = \int \frac{dz^- \, d^2 \bm z_\perp}{2 (2\pi)^3} \, e^{i k \cdot z} \,
 \langle p^{\prime} \, | \, \bar{\psi}(-\tfrac{1}{2}z) \,
 \Gamma \, {\cal W} \,
 \psi(\tfrac{1}{2}z) \, | \, p \rangle\, \Big|_{z^+ = 0} \,.
 \label{eq:gcol}
\end{align}
where $W(P, x, \bm k_\perp,\Delta)$ is the generalized parton correlation function (GPCF) of the pion. For completeness we also write down the decomposition of the GPCF to other twist-2 GTMDs:
\begin{align}
W^{[\gamma^+]} & =  F_{1,1} \,, \vphantom{\frac{1}{1}}
\label{eq:f11} \\
W^{[i\sigma^{j+}\gamma_5]} & =
 -\frac{i\varepsilon_\perp^{ij} \bm k_\perp^i}{M_\pi} \, H_{1,1}
-\frac{i\varepsilon_\perp^{ij} \bm \Delta_\perp^i}{M_\pi} \, H_{1,2} \,,
\label{eq:h1}
\end{align}
Note that unlike $F_1$ and $H_1$ which are the GPD-limit of more general GTMDs by $\bm k_\perp$ integral, there is no corresponding GPD for the GTMD $G_{1,1}$ since it is $\bm k_\perp$-odd. Therefore, the relation Eq.~(\ref{eq:cqgtmd}) provides another expression for $C_z^q$ from a more general structure of the parton correlation.

\section{Model results of the spin-orbit correlation of the pion meson}

In the previous section, we present two different expressions for the quark spin-orbit correlation $C_z^q$ of the pion meson.
One is in terms of GPDs (Eq.~(\ref{eq:1})), the other is in terms of GTMD (Eq.~(\ref{eq:cqgtmd})). In this section, we will provide the model results for $C_z^{q/\pi}$ using these two relations.
We note that in the case of nucleon, the light-front constituent quark model and the light-front chiral quark-soliton model~\cite{Lorce:2011dv} were applied to calculate the quark spin-orbit correlation numerically.
Here, we will provide the analytic result as well numerical result for $C_z^{q/\pi}$ using a light-cone quark model for the pion meson.
The light-cone formalism has been widely used in the parton distribution functions of nucleons and mesons~\cite{Lepage:1979za,Bacchetta:2008af}, and the overlap representation has also been used to study various form factors of the nucleon~\cite{Brodsky:2000ii} and the pion~\cite{Xiao:2003wf}, anomalous magnetic moment of the nucleon~\cite{Lu:2006kt} as well as GPDs~\cite{Brodsky:2000xy}. The reliability of this model is beyond doubt, and the resulting predicted results agree well with the experiments.

In Ref.~\cite{Ma:2018ysi}, the light-cone quark model~\cite{Xiao:2003wf} was applied to calculate the GTMDs of the pion meson, within the overlap representation for the GPCFs.
In this model, the light-cone wave function of
the minimal Fock states $\psi(x,\bm{k}_\perp,\lambda_q,\lambda_{\bar{q}})$ of these wave functions have been derived in Ref.~\cite{Xiao:2003wf} by considering the relativistic effect of quarks~\cite{Melosh:1974cu,Ma:1991xq}:
\begin{align}
\psi(x,\bm{k}_\perp,+,-)&=+\frac{m_q}{\sqrt{2(m_q^2+\bm{k}^2_\perp)}}\phi_\pi  \quad  (l^z=0),\nonumber\\
\psi(x,\bm{k}_\perp,-,+)&=-\frac{m_q}{\sqrt{2(m_q^2+\bm{k}^2_\perp)}}\phi_\pi  \quad  (l^z=0),\nonumber\\
\psi(x,\bm{k}_\perp,+,+)&=-\frac{k_{\perp1}-ik_{\perp2}}{\sqrt{2(m_q^2+\bm{k}^2_\perp)}}\phi_\pi  \quad  (l^z=-1),\nonumber\\
\psi(x,\bm{k}_\perp,-,-)&=-\frac{k_{\perp1}+ik_{\perp2}}{\sqrt{2(m_q^2+\bm{k}^2_\perp)}}\phi_\pi  \quad  (l^z=+1),
\label{eq:wavefunction}
\end{align}
where $+,-$ denotes the helicity of the quark and antiquark, and
\begin{align}
\phi_\pi(x,\bm{k}_\perp)=A\ exp[-\frac{1}{8\beta^2}\frac{\bm{k}_\perp^2+m_q^2}{x(1-x)}].
\end{align}
As shown in Ref.~\cite{Xiao:2003wf}, within the wave functions in Eq.~(\ref{eq:wavefunction}), the light-cone model can describe the transition form factor of the pion meson fairly well.

In this work, we adopt the model results directly from Ref.~\cite{Ma:2018ysi}, in which $F_{1,1}$, $H_{1,1}$, $H_{1,2}$ and $G_{1,1}$ have the expressions:
\begin{align}
F_{1,1}^{q/\pi}&=C(2\bm{k}^2_\perp-\frac{(1-x)^2}{1-\xi^2/4}\frac{\bm\Delta^2_\perp}{2}
-\frac{\xi(1-x)}{1-\xi^2/4}\bm\Delta_\perp\cdot\bm{k}_\perp+2m_q^2)\\
&\times exp(-\frac{(2x(1+\xi^2/4)-\xi^2)(\bm{k}^2_\perp+m_q^2)+x(1-x)^2\Delta^2_\perp/2
-\xi(1-x)^2\bm{k}_\perp\cdot\bm\Delta_\perp}{8\beta^2(x^2-\xi^2/4)(1-x)})\label{eq:f}\\
G_{1,1}^{q/\pi} & =
-C{2(1-x)M_\pi^2\over 1-\xi^2/4}\exp\left( (2x(1+\xi^2/4)-\xi^2)(\bm k_\perp^2 + m_q^2) +x(1-x)^2\bm \Delta_\perp^2/2-\xi(1-x)^2 \bm k_\perp \cdot \bm\Delta_\perp \over 8\beta^2(x^2-\xi^2/4)(1-x)\right), \\
H_{1,1}^{q/\pi}&=0\\
H_{1,2}^{q/\pi}&=C\frac{2(1-x)m_q\,M_\pi}{1-\xi^2/4}exp(-\frac{(2x(1+\xi^2/4)-\xi^2)(\bm{k}^2_\perp+m_q^2)
+x(1-x)^2\Delta^2_\perp/2-\xi(1-x)^2\bm{k}_\perp\cdot\bm\Delta_\perp}{8\beta^2(x^2-\xi^2/4)(1-x)}),
\label{eq:ff}
\end{align}
for the valence quark, where
\begin{align}
C=\frac{A^2}{32\pi^3B_+B_-}
\end{align}
with
\begin{align}
B_+=\sqrt{\left(\bm{k}_\perp+\frac{1-x}{1+\xi/2}\frac{\Delta^2_\perp}{2}\right)^2+m_q^2}\\
B_-=\sqrt{\left(\bm{k}_\perp-\frac{1-x}{1-\xi/2}\frac{\Delta^2_\perp}{2}\right)^2+m_q^2}.
\end{align}

The $\bm k_\perp$-even GTMDs can be reduced to GPDs after integrating over $\bm k_\perp$:
\begin{align}
F_{1}^{q/\pi}(x,\xi, t)& =\int d^2\bm{k}_\perp F_{1,1},\\
H_{1}^{q/\pi}(x,\xi, t) &=\int d^2\bm k_\perp \left({ \bm k_\perp \cdot \Delta_\perp \over\bm \Delta_\perp^2}H_{1,1}+H_{1,2}\right).
\end{align}
Thus, using Eq.~(\ref{eq:1}), in the GPD approach for the spin-orbit correlation,
\begin{align}
C^{q/\pi}_z\big{|}_{\textrm{GPD}} &= \int dx \left(\frac{m_q}{M_\pi} H^{\Delta}_1(x,0,0)- \frac{1}{2} F_1(x,0,0)\right)\nonumber\\
      &= A^2\int dx d^2\bm k_\perp{2(1-x)m_q^2-(\bm k_\perp^2+m_q^2)\over 32\pi^3(\bm k_\perp^2+m_q^2)} \exp\left(-{\bm k_\perp^2+m_q^2\over 4\beta^2 x(1-x)}\right) \\
      &=\int d x C^{q/\pi}_z(x) \nonumber
\label{eq:modelGPD}
\end{align}
In the above equation, we have used $C^{q/\pi}_z(x)$ to denote the integrand.

Since the integration over $x$ satisfies the relation:
\begin{align}
\int dx\ d^2\bm{k}_\perp \,2(1-x) \,\exp\left(-{\bm k_\perp^2+m_q^2\over 4\beta^2 x(1-x)}\right)  = \int dx\ d^2\bm{k}_\perp \, \exp\left(-{\bm k_\perp^2+m_q^2\over 4\beta^2 x(1-x)}\right),
\end{align}
Eq.~(\ref{eq:modelGPD}) can be rewritten as
\begin{align}
C^{q/\pi}_z\big{|}_{\textrm{GPD}}& = A^2\int dx d^2\bm k_\perp {-\bm k_\perp^2\over 32\pi^3(\bm k_\perp^2+m_q^2)} \exp\left(-{\bm k_\perp^2+m_q^2\over 4\beta^2 x(1-x)}\right) \nonumber\\
\end{align}
after integrating over $\bm k_\perp$, the correlation has the form
\begin{align}
C^{q/\pi}_z\big{|}_{\textrm{GPD}} &= -{A^2\over 32\pi^2}\int dx \left(\beta^2 x(1-x)\exp\left(-{m_q^2\over 4\beta^2 x(1-x)}\right)-m_q^2\Gamma\left[0,{m_q^2\over 4\beta^2 x(1-x)}\right]\right), \label{eq:intx}
\end{align}
where $\Gamma[0,x]$ is the zero-th order incomplete $\Gamma$ function
\begin{align}
\Gamma[0,x] = \int_x^\infty {d t\over t} \, e^{-t}.
\end{align}
The integral in over $x$ can be performed numerically.

On the other hand, as shown in Eq.~(\ref{eq:cqgtmd}), $C^{q/\pi}_z$ can be also calculated from the GTMD $G_{1,1}$ directly:
\begin{align}
C^{q/\pi}_{z}\big{|}_{\textrm{GTMD}} &= \int dx d^2\bm k_{\perp} \frac{\bm{k}^2_{\perp}}{M^2} G_{1,}^{\pi}(x,0,\bm{k}^2_{\perp},0,0)\nonumber\\
      &=- A^2\int dx d^2\bm k_\perp{(1-x)\bm k_\perp^2\over 16\pi^3(\bm k_\perp^2+m_q^2)} \exp(-{\bm k_\perp^2+m_q^2\over 4\beta^2 x(1-x)})\nonumber\\
      &=-{A^2\over 32\pi^2}\int dx \left(\beta^2 x(1-x) \exp\left(-{m_q^2\over 4\beta^2 x(1-x)}\right)-m_q^2\Gamma\left[0,{m_q^2\over 4\beta^2 x(1-x)}\right]\right) \nonumber\\
      &=\int d x C^{q/\pi}_z(x).
\label{eq:cqgtmdintx}
\end{align}
We find that the above expression is the same as that in Eq.~(\ref{eq:intx}).
Thus, within the light-cone quark model, we find that the GPD approach and the GTMD approach for the correlation indeed can lead to the same results for $C^{q/\pi}_z$.

In the following, we can obtain the numerical results for $C_z^{q/\pi}$ by adopting the values for the parameters and performing the $x$ integral in Eqs.~(\ref{eq:intx}) or (\ref{eq:cqgtmdintx}).
We follow the choice in Refs.~\cite{Wang:2017onm} for the parameter values:
$ A=31.30\textrm{GeV}^{-1}\, \beta=0.41\,\textrm{GeV}, \,m=0.2\,\textrm{GeV}$.
The numerical result of the quark spin-orbit correlation inside the pion meson in the light-cone quark model is
\begin{align}
C_z^{q/\pi}=-0.32
\end{align}
where $q$ denotes the valence quarks inside pion (e.g., $u$ and $\bar{d}$ in $\pi^+$).
Similar to the case in the nucleon~\cite{Lorce:2014mxa}, the sign of this correlation is negative, which means that the quark longitudinal spin and the quark OAM tend to be anti-correlated inside the pion meson.
The absolute value is smaller than those of the nucleon  ($C_z^{u/n}\approx -0.9$ and $C_z^{d/n}\approx -0.53$~\cite{Lorce:2014mxa}), implicating a weaker correlation in the pion meson than that in the nucleon.

\begin{figure}
  \centering
  \includegraphics[width=0.52\columnwidth]{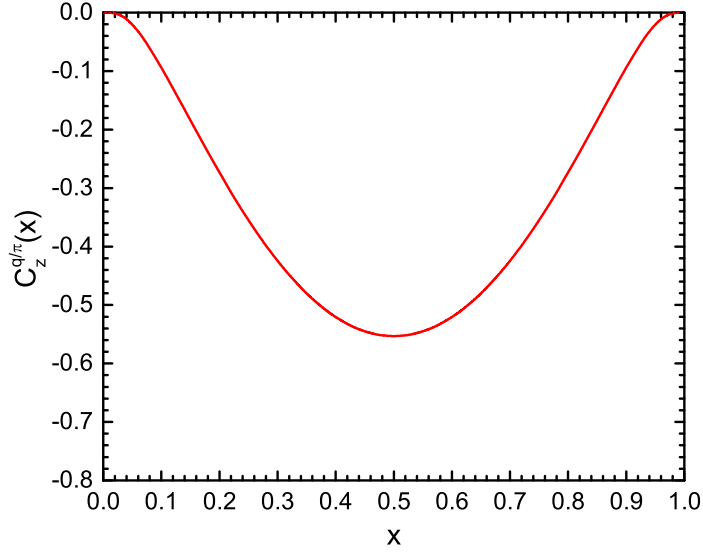}
  \caption{The $x$-dependence of the quark longitudinal spin-orbit correlations $C_z^{q/\pi}(x)$ in the pion meson.}
  \label{fig:czx}
\end{figure}

In order to show the contribution to the quark spin-orbit correlation in the different $x$ region, we
keep $C_z^{q/\pi}$ unintegrated. That is, we calculate $C_z^{q/\pi}(x)$ appearing in Eq.~(\ref{eq:intx}) or Eq.~(\ref{eq:cqgtmdintx}) and show the plot vs $x$ in Fig.~\ref{fig:czx}. We find that in our model the largest contribution comes from the region $x$ around 0.5.

In the following, we also explore the contribution in different region of transverse momentum.
Before doing this, we would like to point out that there is another method to calculate the spin-orbit correlation directly from the wave function of the pion meson, instead of GPD or GTMD.
As $C_z$ is the difference of the orbital angular momenta from left-hand and right-hand quarks.
Using the wave functions in Eq.~(\ref{eq:wavefunction}), $C^{q/\pi}$ can be also expressed as
\begin{align}
C_z^{q/\pi}\big{|}_{\textrm{JM}}&=\hat{L}_z^{q_R}-\hat{L}_z^{q_L} \nonumber\\
&={1\over 16 \pi^3}\int  dx \int d^2 k_\perp \left [(-|\psi(x,\bm{k}_\perp,+,+|^2)-(+|\psi(x,\bm{k}_\perp,-,-)|^2) \right]\nonumber\\
&=- A^2\int dx d^2\bm k_\perp{(1-x)\bm k_\perp^2\over 16\pi^3(\bm k_\perp^2+m_q^2)} \exp\left(-{\bm k_\perp^2+m_q^2\over 4\beta^2 x(1-x)}\right).
\end{align}
We find that the results consistent with the result from the GTMD approach.
We point out that this method is similar to the Jaffe-Manohar approach for the quark OAM~\cite{Jaffe:1989jz}.
We also comment that in this approach, the quark OAM inside the pion meson vanishes,
as
\begin{align}
L_z^{q/\pi}\big{|}_{\textrm{JM}}=\hat{L}_z^{q_R}+\hat{L}_z^{q_L}&={1\over 16 \pi^3}\int  dx \int d^2 k_\perp \left [(-|\psi(x,\bm{k}_\perp,+,+|^2)+(+|\psi(x,\bm{k}_\perp,-,-)|^2) \right] .
\end{align}
Using the wave function in Eq.~\ref{eq:wavefunction}, we find that $L_z^{q/\pi}$ is zero.
It comes from the cancelation of the contributions from the left-handed and right-handed quark.

\begin{figure}
  \centering
  \includegraphics[width=0.48\columnwidth]{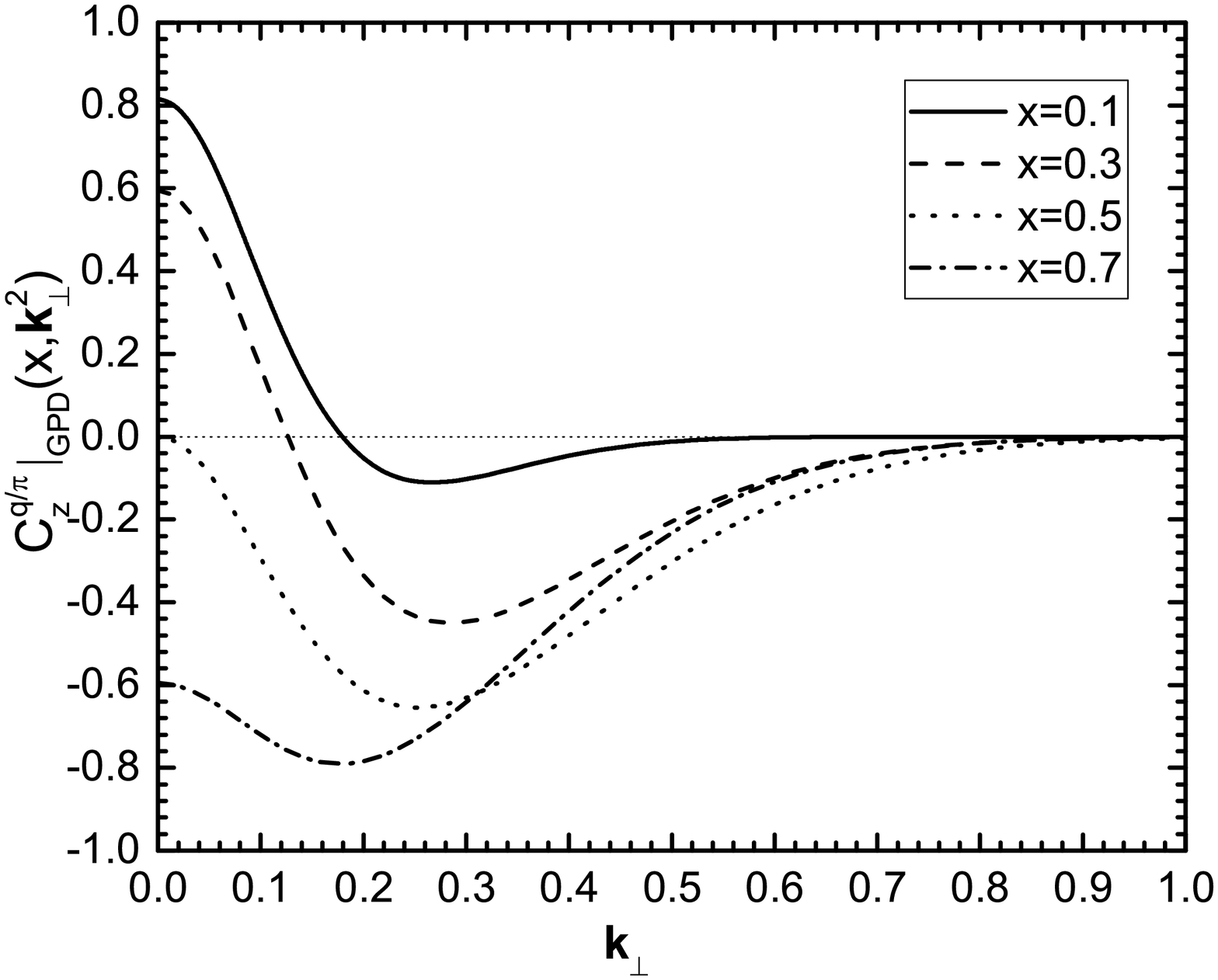}
  \includegraphics[width=0.48\columnwidth]{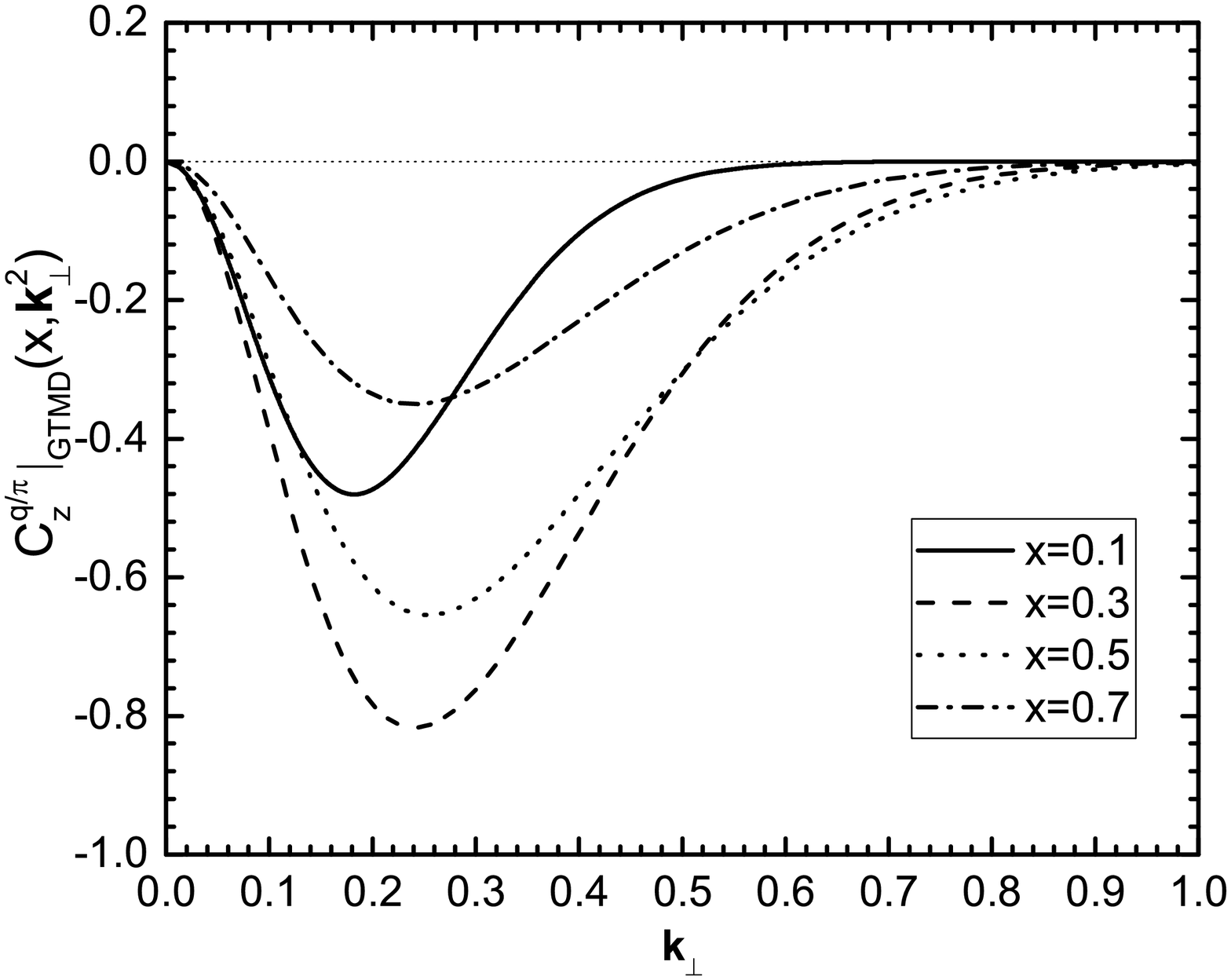}
  \caption{Left panel: The $k_\perp$-dependence of the unintegrated quark correlation $C_z^{q/\pi}(x,\bm{k}_\perp^2)$ from the GPD approach at $x=$0.1, 0.3, 0.5 and 0.7 respectively.
   Right panel: Similar to the left panel, but for that from the GTMD approach.}
  \label{fig:cg}
\end{figure}

In Fig.~\ref{fig:cg}, we plot the $\bm k_\perp$ dependence of $C_z(x,\bm k_\perp)$, which is $C_z$  keeping both $x$ and $\bm k_\perp$ unintegrated.
The left panel shows $C_z(x,\bm k_\perp)$ at $x=0.1$, 0.3, 0.5 and 0.7 from the GPD approach , while the right panel shows that from the GTMD approach.
Our results show that, although $C_z(x)$ is the same in the two approaches, the $\bm k_\perp$-dependence of $C_z(x,\bm k_\perp)$ can be actually different. In the GTMD approach, $C_z(x,\bm k_\perp)$ is negative in the whole reigon, while in the GPD approach, $C_z(x,\bm k_\perp)$ is positive in the small $k_\perp$ when $x$ is not large.

\section{Conclusions}

We studied the correlation between the longitudinal spin and the quark orbit motion of valence quarks inside the pion meson. We started from the parity-odd partner of the quark energy-momentum tensor operator $\hat{T}_{q5}^{\mu\nu}$ and decompose it into form factors, among which the quark spin orbit correlation is determined by the form factor $C_z^q$. We provided two expressions for this correlation $C_z^q$. One is in terms of the GPDs of the pion meson, from which the expectation value of the correlation is given by the combination of the first-$x$ moments of $F_1^{q/\pi}(x,\xi,t)$ and $H_1^{q/\pi}(x,\xi,t)$ at $\xi=0, t=0 \,\textrm{GeV}^2$. The other is in terms of the GTMD $G_{1,1}$.
Using the overlap representation for the pion GPDs and GTMDs derived from a light-cone quark model, we then calculated the analytic result of $C_z^{q/\pi}$.
We found that the result from the GPD approach are the same as that from the GTMD approach.
This verifies from the model aspect that these two approach can be used to access the spin-orbit correlation.
In addition, the numerical result of the quark spin-orbit correlation in the model was calculated as $C_z^{q/\pi}=-0.32$. Similar to the case of the nucleon, the negative sign indicates that the quark spin and OAM is tend to be anti-correlated.
We also present the results for the $x$-dependence and the $k_\perp$-dependence of the longitudinal spin-orbit correlation, that is, the unintegrated $C_z^{q/\pi}$. Our study about the quark longitudinal spin provides a new information for the spin correlation inside the pion meson, further experimental measurement are needed to accurately determine this correlation.

\section*{Acknowledgements}
This work is partially supported by the National Natural Science Foundation of China under grant number 11575043.

\end{document}